\documentstyle[aps,12pt,twocolumn]{revtex}
\begin{document}
\title{Optical Realization of Quantum Gambling Machine}
\author{Yong-Sheng Zhang, Chuan-Feng Li, Wan-Li Li, Yun-Feng Huang, and Guang-Can Guo%
\thanks{%
Electronic address: gcguo@ustc.edu.cn}}
\address{Laboratory of Quantum Communication and Quantum Computation and Department\\
of Physics, University of Science and Technology of China, Hefei 230026, P.\\
R. China}
\maketitle

\begin{abstract}
\baselineskip12ptQuantum gambling --- a secure remote two-party protocol
which has no classical counterpart --- is demonstrated through optical
approach. A photon is prepared by Alice in a superposition state of two
potential paths. Then one path leads to Bob and is split into two parts. The
security is confirmed by quantum interference between Alice's path and one
part of Bob's path. It is shown that a practical quantum gambling machine
can be feasible by this way.
\end{abstract}

\baselineskip12ptAs a kind of game, gambling plays an important role in the
society and nature which are full of conflict, competition and cooperation.
Up to now, game theory has been investigated with mathematical methods [1]
and applied to study economy, psychology, ecology, biology and many other
fields [2, 3].

One might wonder why games like gambling can have anything to do with
quantum physics. After all, game theory is about numbers that entities are
efficiently acting to maximize or minimize. However, if linear
superpositions of the actions are permitted, games will be generalized into
quantum domain [4, 5]. Quantizing games may be interesting in several fields
[4], such as foundation of game theory, games of survival and quantum
communication [6]. Moreover, quantum mechanics may assure the fairness in
remote gambling [7].

In this letter, we present a quantum gambling machine composed of optical
elements.

We may firstly investigate the simplest classical gambling machine: one
particle and two boxes $A$ and $B$. During a game, the casino (Alice) stores
the particle in $A$ or $B$ randomly, then the player (Bob) guesses which box
the particle is in. For the two parties do not trust each other, even a
third party, a remote classical gambling is impossible. Whereas in the
quantum domain, Alice may prepare the particle in a superposition state of $%
\left| a\right\rangle $ (the particle in $A$) and $\left| b\right\rangle $
(the particle in $B$). If she generate the equal superposition state 
\begin{equation}
\left| \Psi _0\right\rangle =\frac 1{\sqrt{2}}\left( \left| a\right\rangle
+\left| b\right\rangle \right)  \eqnum{1}
\end{equation}
and a prescribed box (e.g. $B$) is sent to Bob, a remote fair gambling may
be carried out. For simplicity, the bet in a single game is taken to be one
coin. If Bob finds the particle in box $B$ (state $\left| b\right\rangle $),
he wins one coin, otherwise he loses the bet. Obviously, the probability for
Bob to win is exactly $50\%$. Moreover, Bob cannot cheat by claiming that he
found the particle when he did not, for Alice can verify by opening box $A$.

In order to decrease the probability for the particle in box $B$, Alice may
prepare a biased superposition state (she gets no advantage using an ancilla
or other complex strategy [7]) 
\begin{equation}
\left| \Psi _0^{\prime }\right\rangle =\sqrt{\frac 12+\epsilon }\left|
a\right\rangle +\sqrt{\frac 12-\epsilon }\left| b\right\rangle  \eqnum{2}
\end{equation}
instead of $\left| \Psi _0\right\rangle $, where $\epsilon $ is the
preparation parameter, with $0\leq \epsilon \leq \frac 12$. However, the
quantum principle assures that Bob has a chance to find out the difference
and win her $R$ coins, which is the punishment the two parties agree on
before the game.

Bob's strategy is to split out part of the state $\left| b\right\rangle $
and convert it to state $\left| b^{\prime }\right\rangle $ by performing a
unitary operation, {\it i.e.}, 
\begin{equation}
\left| b\right\rangle \rightarrow \sqrt{1-\eta }\left| b\right\rangle +\sqrt{%
\eta }\left| b^{\prime }\right\rangle ,  \eqnum{3}
\end{equation}
Where $\left| b^{\prime }\right\rangle $ is orthogonal to $\left|
a\right\rangle $ and $\left| b\right\rangle $ and $\eta $ is the splitting
parameter. After the splitting, if Bob does not find the particle in box $B$%
, Alice will send box $A$ to him for verification. In this case the state of
the particle is reduced to $\left| \phi _a\right\rangle =\sqrt{\frac 1{%
1+\eta }}\left( \left| a\right\rangle +\sqrt{\eta }\left| b^{\prime
}\right\rangle \right) $, if Alice prepare the particle in the equal
superposition state $\left| \Psi _0\right\rangle $. Therefore, the
verification of Bob is to measure the particle under the basis $\left| \phi
_a\right\rangle $ and its orthogonal basis $\left| \phi _b\right\rangle $.
If Alice prepare the biased superposition state $\left| \Psi _0^{\prime
}\right\rangle $, he may find the particle in state $\left| \phi
_b\right\rangle $ with a certain probability and win $R$ coins.

There exists an equilibrium for the two parties in this protocol [7]. Alice
can ensure her expected gains no less than zero by preparing the equally
distributed state $\left| \Psi _0\right\rangle $. Bob can ensure his
expected gains no less than a particular value only depending on $R${\it \ }%
by selecting an optimal splitting parameter $\eta =\tilde \eta \left(
R\right) $. In fact, this protocol is a zero-sum game, and the strategies of
Alice and Bob are represented by different choices of $\epsilon $ and $\eta $%
, respectively.

In the experiment, a linear-polarized photon is employed as the particle.
Similar to the simulation of quantum logic [8], two potential paths of the
photon may serve as boxes $A$ and $B$. $\left| b^{\prime }\right\rangle $
are distinguished from $\left| a\right\rangle $ and $\left| b\right\rangle $
by the polarization of the photon.

\begin{center}
{\bf Figure 1}
\end{center}

The setup of the optical quantum gambling machine is shown in Figure 1. A
virtue of this machine is that all the detections are carried out
automatically by the machine, which may help to eliminate the classical
communication between the parties and prevent their cheating.

Initially, the photon is generated in a definite linear polarization state
(such as vertical $\left| V\right\rangle $ or horizontal $\left|
H\right\rangle $) by a polarizer. Then the state is transferred to a
superposition state of $\left| V\right\rangle $ and $\left| H\right\rangle $
with half waveplate (HWP) $a$ according to the preparation parameter $%
\epsilon $ chosen by Alice. The preparation is accomplished by swapping the
location and polarization states of the photon with polarizing beamsplitter
(PBS) $1$ and the fixed HWP $\sigma _x$. After the state swapping, the
polarization is horizontal while the location is prepared in the required
state $\left| \Psi _0^{\prime }\right\rangle $.

Bob's splitting is realized by adjusting the HWP $b_1$ according to the
parameter $\eta $ he selects. Then $\left| b^{\prime }\right\rangle $ (split
out by Bob) is separated from $\left| b\right\rangle $ via PBS $2$ and
superposed with $\left| a\right\rangle $ via PBS $3$. The verification is
implemented with HWP $b_1$ and PBS $4$. HWP $b_1$ is adjusted according to $%
\eta $ so as to assure that $\left| \phi _a\right\rangle $ and $\left| \phi
_b\right\rangle $ are transmitted and reflected by PBS $4$ respectively. In
order to obtain the result of the gambling, three detectors $D_1$, $D_2$ and 
$D_3$ are adopted to detect the photon in the state $\left| b\right\rangle $%
, $\left| \phi _a\right\rangle $ and $\left| \phi _b\right\rangle $,
respectively.

A single game of gambling with this machine is described as follows. After
Bob put in his bet --- one coin, the machine will inform Alice and Bob to
select the parameter $\epsilon $ (adjusting HWP $a$) and $\eta $ (adjusting
HWP $b_1$ and $b_2$ simultaneously). Then a photon is generated from the
polarizer and distributed to three parts. If the detector $D_1$ or $D_3$
responds, Bob win one or $R$ coins; if $D_2$ responds, Bob loses the bet
(then the bet will be conserved for Alice automatically).

To demonstrate the performance of the optical gambling machine, a beam
(composed of independent identical photons) is generated instead of a single
photon during the experiment, namely, a well polarized He-Ne laser (3mW at
632.8nm) is utilized as the light source. The results are shown in Figure 2,
where $P_1$ and $P_3$ denote the probabilities that Bob win one and $R$
coins, $P_2$ denotes the probability that Bob lose the bet. The
probabilities are determined by the relative light intensities measured by
the three detectors.

\begin{center}
{\bf Figure 2}
\end{center}

In order to illustrate Bob's strategies, we suppose that Alice and Bob agree
on $R=5$ at the beginning of the gambling. The expected gains of Bob are
shown in Figure 3. Obviously, there exists an optimal splitting parameter $%
\tilde \eta \left( 5\right) \doteq 0.27$ to assure his expected gains no
less than a particular value despite Alice's choice.

\begin{center}
{\bf Figure 3 }
\end{center}

Optical approach has many advantages. By making use of two different freedom
degrees of the photon (location and polarization), an optical quantum
gambling machine may be realized conveniently with several HWPs, PBSes and
detectors. Particularly, the decoherence of all-optical system is relatively
low [9], while the protocol is very sensitive to the errors caused by the
device and environment. As discussed by Goldenberg {\it et al. [7]}, for a
successful realization of quantum gambling, the error rate has to be lower
than $\sqrt{2/R^3}$. Since the error rate in the experiment is only about $%
\frac 1{40}$, a practical quantum gambling may be carried out with this
optical machine under the condition $R<14.4$.

Our experiment has shown that quantum gambling and quantum games have real
physical counterpart, and a practical quantum gambling machine can be
realized with simple optical devices. It can be expected that quantum
mechanics may bring other interesting results in game theory.

This work was supported by the National Natural Science Foundation of China
and the Doctoral Education Fund of the State Education Commission of China.


\begin{references}
\bibitem{}  \baselineskip12ptJ. Von Neumann and O. Morgenstern, {\it The
theory of Games and Economic\ Behavior} (Princeton University Press,
Princeton, 1947).

\bibitem{}  R. B. Myerson, {\it Game Theory: An Analysis of Conflict} (MIT
Press, Cambridge, 1991).

\bibitem{}  R. Axelrod, {\it The Evolution of Cooperation} (Basic Books, New
York, 1984); R. Dawkins, {\it The Selfish Gene} (Oxford University Press,
Oxford, 1976).

\bibitem{}  J. Eisert, M. Wilkens, M. Lewenstein, {\it Phys. Rev. Lett. }%
{\bf 83}, 3077 (1999).

\bibitem{}  D. A. Meyer, {\it Phys. Rev. Lett. }{\bf 82}, 1052 (1999).

\bibitem{}  C. H. Bennett, F. Bessette, G. Brassard, L. Salvail, J. Smonlin, 
{\it J. Crypto.} {\bf 5}, 3 (1992).

\bibitem{}  L. Goldenberg, L. Vaidman, S. Wiesner, {\it Phys. Rev. Lett. }%
{\bf 82}, 3356 (1999).

\bibitem{}  N. J. Cerf, C. Adami, P. G. Kwiat, {\it Phys. Rev. A} {\bf 57},
1447 (1998).

\bibitem{}  P. G. Kwiat, J. R.\ Mitchell, P. D. D. Schwindt, A. G. White,
quant-ph/{\bf 9905086} at http://xxx.lanl.gov.

\baselineskip12ptFigure captions:

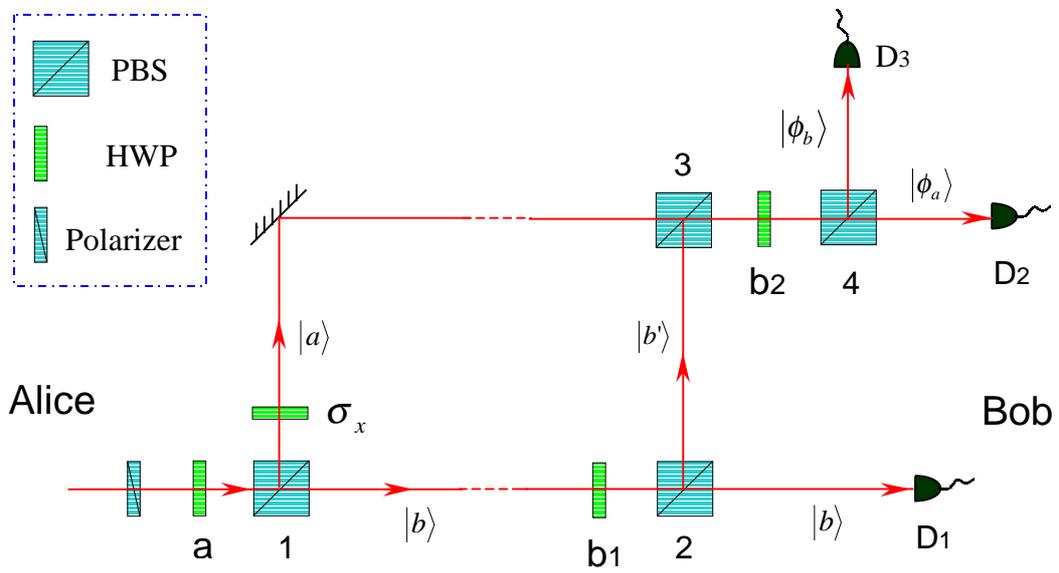
{\bf Figure 1. }The optical setup for quantum gambling machine.

The PBSes transmit the horizontal and reflect the vertical component of the
photons. HWP $\sigma _x$ is fixed at $45^{\circ }$. HWP $a$ is used for
Alice to adjust $\epsilon $ for preparation. HWP $b_1$ is adopted for Bob to
adjust $\eta $ for splitting. HWP $b_2$ is utilized to accomplish the
verification. The phase difference between the two paths from PBS $1$ to PBS 
$3$ are tuned to zero in advance. If the photon is detected by $D_1$ or $D_3$%
, Bob wins one or $R$ coins, respectively. If it is detected by $D_2$, Bob
loses one coin.

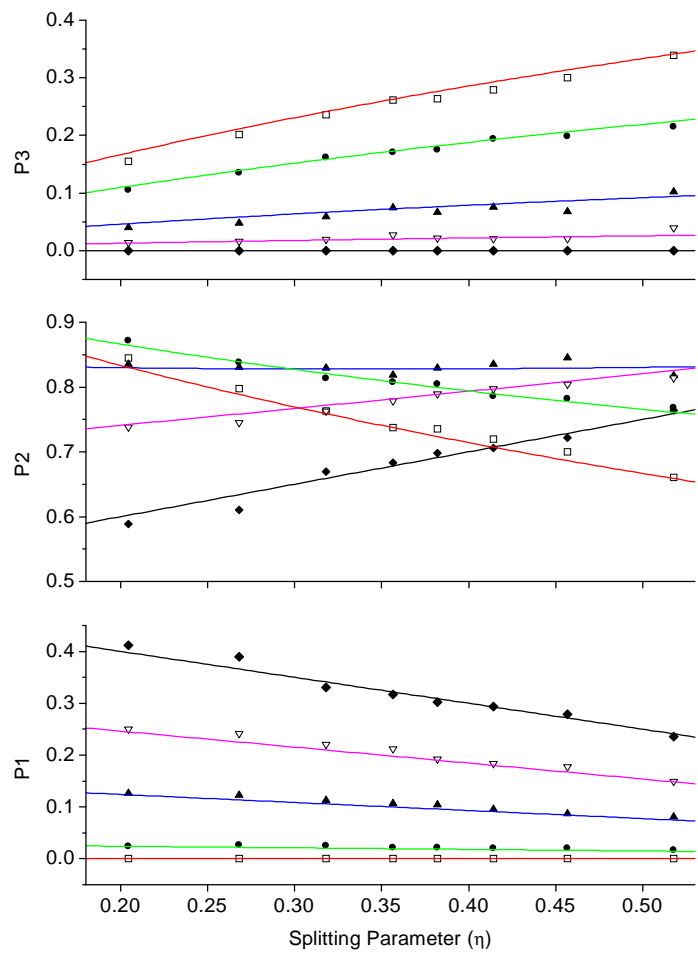
{\bf Figure 2. }Performance of the machine.

$P_1$, $P_2$ and $P_3$ denote the probabilities that Bob wins one, loses one
and wins $R$ coins, respectively. With a certain preparation parameter $%
\epsilon $, the probabilities vary with the splitting parameter $\eta $. The
experimental data are denoted by scattered dots. The solid diamond, open
downtriangle, solid uptriangle, solid circle and open square represent the
cases that $\epsilon =0$, $0.19$, $0.34$, $0.47$ and $0.5$, respectively.
The corresponding lines are theoretical predictions.

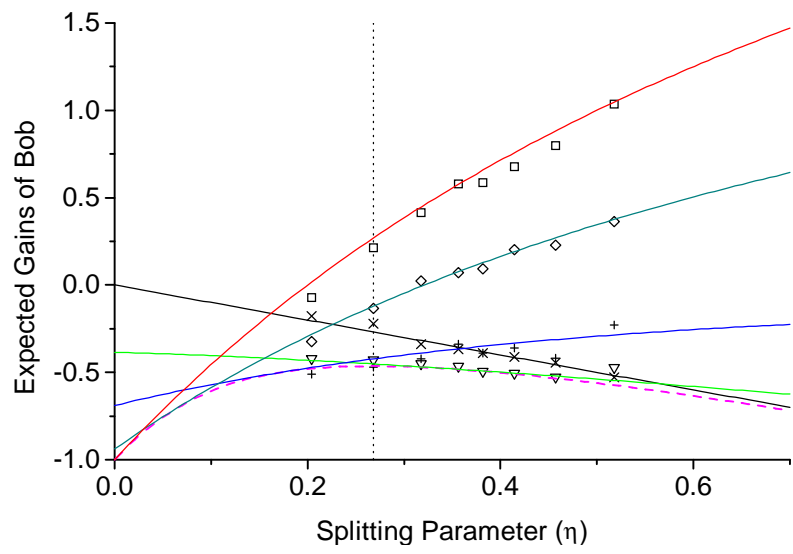
{\bf Figure 3. }Expected gains of Bob varying with $\epsilon $ and $\eta $.

Experimental results are denoted by scattered dots. The cross ($\times $),
downtriangle, cross ($+$), diamond, square represent the case that $\epsilon
=0,$ $0.19,$ $0.34,$ $0.47$ and $0.5$, respectively. The corresponding lines
are theoretical predictions. The lower bound of all possible values is
denoted by the dashed line. It is shown that the optimal parameter $\tilde 
\eta \left( 5\right) \dot =0.27$ because at this value the maximum of the
lower bound is accessed.
\end{references}
\end{document}